\documentclass[modern]{aastex631}

\usepackage{amsmath}
\usepackage{graphicx}
\usepackage[dvipsnames]{xcolor}
\usepackage{xspace}
\usepackage{amsmath}
\usepackage[sort&compress]{natbib}
\usepackage[hang,flushmargin]{footmisc}

\AtBeginDocument{}

\newcommand{\spot}{_\mathrm{spot}}
\newcommand{\fspot}{\ensuremath{f\spot}}
\newcommand{\xspot}{\ensuremath{x\spot}}

\newcommand{\Tc}{\ensuremath{\mathrm{T_{c}}}\xspace}
\newcommand{\Tamb}{\ensuremath{T_\mathrm{amb}}}
\newcommand{\Tspot}{\ensuremath{T_\mathrm{spot}}}
\newcommand{\logrhk}{\ensuremath{\log{\mathrm{R'_{HK}}}}\xspace}

\begin{document}

\title{Star spots mimic the proposed abundance signatures of planet formation}

\author[0000-0003-0174-0564]{Andrew R. Casey}
\affiliation{Center for Computational Astrophysics, Flatiron Institute, 162 Fifth Avenue, New York, NY 10010, USA}
\affiliation{School of Physics and Astronomy, Monash University, VIC 3800, Australia}

\author[0000-0001-9907-7742]{Megan Bedell}
\affiliation{Center for Computational Astrophysics, Flatiron Institute, 162 Fifth Avenue, New York, NY 10010, USA}

\author[0000-0001-7339-5136]{Adam Wheeler}
\affiliation{Center for Computational Astrophysics, Flatiron Institute, 162 Fifth Avenue, New York, NY 10010, USA}

\author[0000-0002-1264-2006]{Julianne Dalcanton}
\affiliation{Center for Computational Astrophysics, Flatiron Institute, 162 Fifth Avenue, New York, NY 10010, USA}

\author[0000-0002-3852-3590]{Lily Zhao}
\affiliation{Center for Computational Astrophysics, Flatiron Institute, 162 Fifth Avenue, New York, NY 10010, USA}

\begin{abstract}
Trends with stellar elemental abundances and condensation temperature have been proposed as a possible tracer of rocky planet formation.
Here we present an alternative explanation, where we show that stellar activity (star spots) can also cause apparent abundance changes in refractory elements preferentially over volatile elements. This effect arises naturally from atomic physics: refractory (rocky; high-\Tc) elements preferentially have low first-ionization potentials, with strong transitions that are most sensitive to the cooler, magnetically-perturbed spot conditions, making any spot-induced change in line strength correlated with \Tc.
We re-examine literature on Sun-like stars in the Solar neighborhood and find that the \logrhk\ stellar activity indicator is correlated with the so-called \Tc\ slope: the gradient of chemical abundances with respect to condensation temperature.
Using 1D LTE \texttt{Korg.jl} simulations of the Sun with spots of varying size and contrast, we show that a wide range of spot configurations can manifest as abundance patterns that mimic a signal of abundance trends with condensation temperature.
Finally, we conduct a line-by-line differential analysis of an ambient Solar spectrum relative to a Solar sunspot umbra to represent an extreme case, where we find that elements with high condensation temperature can show a factor of three (0.5\,dex) increase.
We discuss implications that stellar activity poses for precision spectroscopy, and provide some direction on how this effect might be mitigated.
\end{abstract}

\keywords{Stellar abundances (1577)
---	Stellar photospheres (1237)
--- 	Exoplanet formation (492)
---	Stellar activity (1580)
}

\section{Introduction} \label{sec:introduction}

Detailed chemical abundances of Sun-like stars have become a sensitive diagnostic of the astrophysical processes that shape them: stellar nucleosynthesis, Galactic chemical evolution, and -- as has been proposed -- the formation and subsequent evolution of their planetary systems \citep{Melendez2009}. Through line-by-line differential analysis of high signal-to-noise spectra, the element abundances of a solar twin relative to the Sun can be measured to a precision of $\sim 0.01$\,dex \citep{Melendez2009,Bedell2018}. At this precision a striking pattern has emerged: when compared to Sun-like stars in the Solar neighborhood, the Sun appears deficient in refractory (rocky) elements -- those with a high equilibrium condensation temperature \Tc\ -- relative to volatiles \citep{Melendez2009,Ramirez2009}.\\

Because refractory elements condense out of a cooling protoplanetary disk before volatiles, they are the principal building blocks of rocky bodies. Motivated by this, \citet{Melendez2009} proposed that the Sun's deficit in refractory elements is a result of material being sequestered by terrestrial planets during the formation in the proto-Solar system. Neighboring stars that lack rocky planets would retain this material in their photospheres, and show higher abundances of refractory elements. The gradient of element abundances with respect to the temperature at which 50\% equilibrium condensation occurs (\Tc), colloquially referred to as the `\emph{$T_c$ slope}', has since become the standard statistic used to quantify the effect. The tantalizing possibility of indirectly detecting rocky exoplanets through the chemical abundances of their host star has motivated nearly two decades of effort in precision spectroscopy. \\

Such a signature would be transformative. Measuring element abundances of a single star is orders of magnitude cheaper than the extreme-precision radial velocity or transit programs that otherwise dominate rocky planet detection, and precise chemical abundances are not subject to requiring favorable viewing geometries. Substantial effort has been invested seeking to confirm, refute, or reinterpret \Tc\ trends. In some systems the sign of the \Tc\ slope is consistent with rocky exoplanets detected through radial velocity searches \citep{Melendez2009,Nibauer2021}, while in other systems -- notably stars hosting close-in giant planets -- the \Tc\ slope has the opposite sign \citep{Schuler2011}, prompting a reinterpretation from planet formation to planet \emph{engulfment}, where the ingestion of rocky material might leave an \emph{excess} of refractories in the outer convective envelope \citep{Oh2018,Spina2021,Liu2024}. However, these mechanisms are not without difficulty: engulfment signatures are expected to be rapidly diluted as the convective envelope of the proto-Star deepens, and detailed stellar modeling suggests detections should be rare \citep{Behmard2023a,Behmard2023c}.\\

Several alternative explanations have also been advanced. \citet{Chambers2010} showed quantitatively that the removal of a few Earth masses of rocky material could reproduce the amplitude of the Solar deficit, while \citet{Booth2020} demonstrated that giant planet formation and dust trapping in protoplanetary discs can likewise withhold refractories from the star. Galactic chemical evolution naturally generates age-dependent abundance variations at fixed \Tc, because different elements are enriched on different timescales. The \Tc\ slopes of solar twins have been shown to correlate with stellar age \citep{Adibekyan2014,Nissen2015,Spina2016,Bedell2018}, such that even a sample of nominal solar twins spans a range of ages, leading one to expect \Tc\ slopes without invoking planets. \citet{Rampalli2025} also showed that nucleosynthetic enrichment patterns shaped by Galactic chemical evolution and local interstellar medium inhomogeneities can account for $>90\%$ of the observed star-to-star variation in \Tc\ slope, with the Sun within just $0.5\sigma$ of the expected solar-analog distribution. \\

It is also worth recalling that abundance patterns qualitatively similar to the Solar-neighborhood \Tc\ signal are well-established elsewhere in astrophysics, at amplitudes one to two orders of magnitude larger, and with origins that have nothing to do with terrestrial planet formation. $\lambda$~Boo-type stars -- late-B to early-F dwarfs that are chemically peculiar -- show deficits of refractory elements of up to $\sim 2$\,dex alongside near-solar abundances of C, N, O, and S, producing a pronounced trend of photospheric abundance with \Tc\ \citep{Andrievsky2002}. The favored explanation is selective accretion of circumstellar or interstellar gas from which refractory elements have already condensed onto dust grains \citep{KampPaunzen2002}, and the phenomenon is seen in both young stars and apparently isolated field stars. In the interstellar medium itself, the \emph{gas-phase} abundances of refractory elements are depleted by factors of $10$--$10^2$ relative to volatiles along essentially every sight line studied, through the same sequestration of refractories onto dust grains \citep{SavageSembach1996,Jenkins2009}. In both settings, the same qualitative pattern -- refractories deficient, volatiles near-solar, scaling smoothly with \Tc\ -- arises from grain condensation physics rather than from planetary accretion histories, which demonstrates that the mere existence of a \Tc\ trend in a stellar photosphere is not, by itself, evidence for planet formation or engulfment.\\

Set against these astrophysical contexts, the Solar-neighborhood signal is remarkably small. The \Tc\ slopes reported for solar twins are of order $10^{-4}\,\mathrm{dex\,K}^{-1}$, corresponding to only $\sim 0.1$\,dex spread across the full refractory-to-volatile range of \Tc. This is only about an order of magnitude above the $\sim 0.01$\,dex per-element precision achievable by the most careful line-by-line differential analyses of solar twins \citep{Melendez2009,Nissen2015,Bedell2018}, and the slopes reported for the least-active stars in the \citet{Bedell2018} sample are formally consistent with zero. At this amplitude, the interpretation of any individual \Tc\ trend depends sensitively on whether systematic effects of the same size -- galactic chemical evolution, non-LTE and 3D corrections, model-atmosphere interpolation errors, microturbulence assumptions, continuum placement, or (as we argue below) stellar activity -- have been fully controlled in the differential comparison.\\

A less-explored possibility is that \Tc\ trends are not genuine abundance differences at all, but artefacts of analyzing an inhomogeneous (e.g., spotted) photosphere with a homogeneous model atmosphere: physics that is missing from the model is absorbed into the inferred abundances. \citet{Ramirez2009} noted that the abundance differences between their solar analogues also correlated with first ionization potential (FIP) -- which in turn correlates with \Tc\ -- as would be expected if the abundance differences arose from unmodeled photospheric inhomogeneity rather than planet formation. This scenario was nonetheless set aside in favor of a planet-formation interpretation, and has received comparatively little attention since. \citet{Spina2020} subsequently showed that stellar magnetic activity can introduce substantial systematics in high-precision abundance work on Sun-like stars, because cool, magnetized surface features (spots and plage) alter line strengths in ways that correlate with excitation potential, FIP, and Zeeman sensitivity. That work did not, however, make the explicit connection between these activity-induced systematics and the \Tc\ slope, nor did it quantify whether spot configurations plausible for solar-type stars can reproduce the amplitude and sign of the \Tc\ trends attributed to planet formation or engulfment. Establishing that connection is the central aim of this work.\\

Similarly, various works have shown that stellar activity can cause apparent changes in stellar parameters and overall metallicity \citep{Cao2022,Wilson2023}.
Independently\footnote{Most of the work for this paper was completed in 2023-2024. Time is strange.}, \citet{Yu2025} find that within co-natal co-moving stellar pairs, the star more depleted in refractory elements is also more magnetically active, with the correlation strengthening with condensation temperature and absent for volatile elements. They interpret this as an imprint of planet formation, arguing that refractory sequestration and enhanced magnetic activity share a common origin in pre-main-sequence disc--star coupling. We argue below that the same correlation is also a natural consequence of spot-induced spectroscopic bias, and that the two scenarios are observationally degenerate without additional constraints.
This connection is not coincidental: the refractory elements that condense at high \Tc\ -- iron-peak and $\alpha$-process metals such as Fe, Ni, Mg, Al, Ca, and Ti -- preferentially have low FIPs because the same atomic-structure properties (loosely-bound valence electrons in metals) drive both ease of ionization and ease of grain incorporation. The strong, low-excitation resonance lines of these low-FIP species are precisely the lines whose source functions and equivalent widths respond most strongly to the cooler temperatures and stronger magnetic fields characteristic of spotted regions, so a spot-induced shift in line strength produces a smoothly-varying signal in \Tc. \\

In this work we investigate whether stellar activity, on its own, can produce chemical abundance patterns that mimic the reported \Tc\ signature of planet formation and engulfment. We present three complementary experiments. In Section~\ref{sec:solar-like-stars} we re-measure the chromospheric activity indicator \logrhk\ from the HARPS spectra used by \citet{Bedell2018}, and show that the reported \Tc\ slopes of solar twins are correlated with stellar activity. In Section~\ref{sec:simulated-data} we simulate a spotted Sun with a two-parameter model of spot filling fraction and contrast, and show that plausible spot configurations reproduce both the amplitude and either sign of the \Tc\ slopes reported in the literature. In Section~\ref{sec:observed-solar-spectra} we perform a line-by-line differential analysis of a quiet Solar spectrum against a sunspot umbral spectrum, representing an extreme case, and find that estimated refractory abundances can be up to three times higher (0.5\,dex) than volatiles. We discuss the implications for high-precision chemical abundance work and mitigation strategies in Section~\ref{sec:discussion}, before concluding in Section~\ref{sec:conclusions}.

\section{Experiments} \label{sec:experiments}

We describe three experiments to explore how stellar activity might impact chemical abundances:
    we test whether stellar activity indicators are correlated with reported slopes of element abundance with condensation temperature (Section~\ref{sec:solar-like-stars});
    we simulate the Sun with spots of varying size and contrast (Section~\ref{sec:simulated-data}); and
    we analyze an ambient (unspotted) Solar spectrum and compare it with a sunspot umbral spectrum (Section~\ref{sec:observed-solar-spectra}).

\subsection{Stellar activity indicators} \label{sec:solar-like-stars}

Stellar activity is often estimated by spectral indices, where spectra are integrated relative to some baseline around strong transitions. Chromospheric activity causes measurable levels of emission from these strong lines, providing an indirect measure of stellar activity. Activity measures have been defined for numerous strong lines, but the Ca~II H \& K lines are usually taken as the most sensitive features for defining a variability index, and the \logrhk\ measure on the Mount Wilson scale \citep{Noyes1984} has emerged as the preferred measure of comparison.\\

We measured \logrhk\ for all 79 Sun-like stars in the Solar neighborhood from \citet{Bedell2018}. This catalog is among the largest homogenous measurements of abundance differences with respect to condensation temperature. Since these stars may have their own cycles where stellar activity varies, we measured \logrhk\ from the \emph{exact same spectra} used to measure abundance gradients with condensation temperature. Specifically, we computed \logrhk\ directly from the co-added HARPS spectra used by \citet{Bedell2018} for abundance analysis. We constructed triangular bandpass filters of 1.09\,\AA\ each around the Ca~II H \& K lines, and rectangular bandpass filters of width 20\,\AA\ centered on 3901.07\,\AA\ and 4001.07\,\AA\ to estimate the continuum. See \citet{Lovis2011} for more details, which includes the calibration used to map the resulting indices onto the Mount Wilson scale.\\

Throughout this paper we define the \Tc\ slope as the gradient of element abundances with respect to the 50\% condensation temperatures \Tc\ of \citet{Lodders2003}. For reference, using the \Tc\ slopes reported by \citet{Bedell2018}, we fit a straight line between the \Tc\ slope and \logrhk\ accounting for the uncertainties in both quantities, which we assume to be uncorrelated. We implemented this model in the \texttt{Stan} probabilistic programming language \citep{stan} and sampled the posterior distribution using a Hamiltonian Monte Carlo sampler. We find the relation (with $1\sigma$ uncertainties)
\begin{equation}
    \frac{\mathrm{d}{[\mathrm{X/H}]}}{\mathrm{d}\Tc} = \left[({3.8} \pm 0.6)\times10^{-4}\,\logrhk + ({1.9 \pm 0.3})\times10^{-3}\right]\,\mathrm{dex\,K}^{-1}\,,
\end{equation}

\begin{figure}
    \centering
    \includegraphics{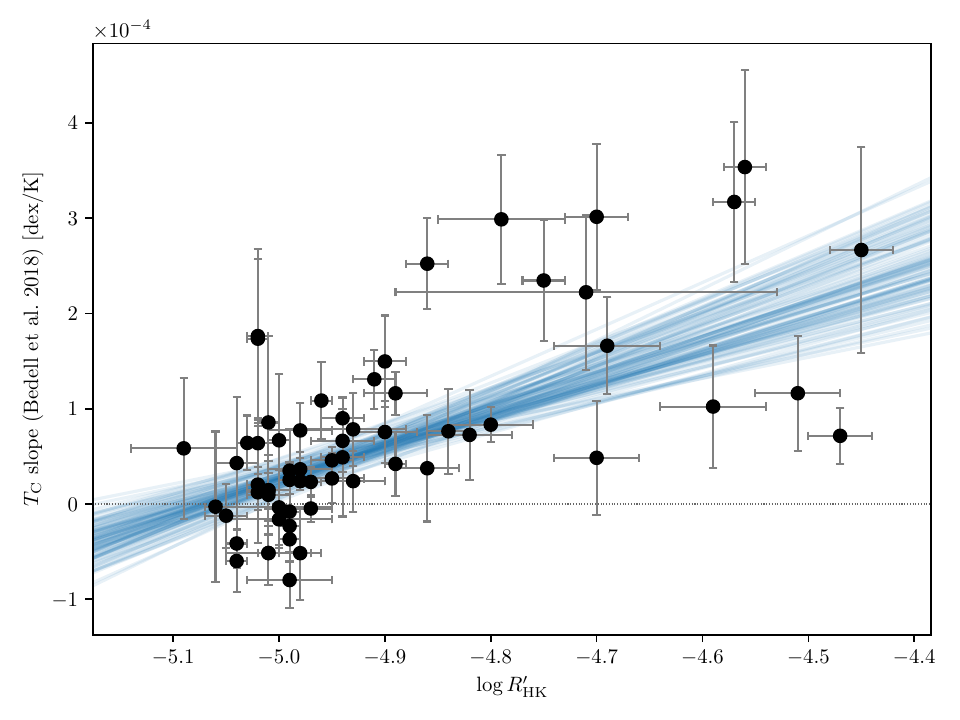}
    \caption{The gradient of element abundances with respect to condensation temperature (\Tc; y-axis) reported by \citet{Bedell2018} as a function of the stellar activity indicator \logrhk. Stars with high stellar activity indicators are more likely to show abundance trends with condensation temperature. \label{fig:bedell}}
\end{figure}

\noindent{}which we illustrate in Figure~\ref{fig:bedell} by taking random draws from our best-fit model and projecting them to the data space. We caution against over-interpreting this fit: the model includes no intrinsic scatter, assumes a linear relation for which there is no physical motivation, and does not account for covariates such as stellar age that correlate with both quantities. We include it only to quantify the trend apparent in Figure~\ref{fig:bedell}.The stars with the lowest \Tc\ slopes are preferentially those with the lowest measures of stellar activity. It would be inappropriate to make causal arguments from these results alone, but Figure~\ref{fig:bedell} does demonstrate that there is a correlation between stellar activity indicators and the slope of element abundances with respect to condensation temperature. An independent corroboration comes from \citet{Yu2025}, who find the same qualitative trend in co-natal co-moving pairs using the Ca~II infrared triplet as an activity indicator: the more refractory-depleted member of each pair is systematically more magnetically active, with the differential signal scaling with \Tc.

\subsection{Simulations of the spotted Sun} \label{sec:simulated-data}

Here we simulate the emergent spectra of the `Sun-as-a-star' with different spot model configurations and evaluate how they impact the inferred chemical abundances.
We adopt a simple spot model represented by just two parameters: the fraction of observable surface area covered by spots ($\fspot$), and the contrast ratio of those spots relative to the ambient surface ($\xspot$).
While a detailed geometric representation of spots would be more realistic (e.g., including a distribution of spot sizes and their latitudes, or decay timescales), we are primarily interested in the integrated spectrum, which is sufficiently represented by our two nuisance parameters. In this model a totally inactive star can have either zero spots ($\fspot = 0$) or no contrast ($\xspot = 1$).\\

We compute emergent spectra for combinations of $\{\xspot, \fspot\}$ trialled on a grid. We take $\fspot$ between $\fspot = 0$ (no spots) and $\fspot = 0.5$ (50\% covered by spots) with a step size of $\Delta\fspot = 0.05$. We adopt these as limits for our simulations, but in reality active stars may have $\fspot$ values that plateau near $\fspot \approx 0.3$ \citep{Cao2022}. We restrict \xspot\ between 0.6 and 1 with $\Delta\xspot = 0.05$. The spot temperature is taken to be $\Tspot = \xspot\Tamb$, and our lower limit on \xspot\ is set by the lower effective-temperature boundary of the model atmosphere grids we use (e.g., $\Tspot = 3466\,\mathrm{K}$ for the Sun when $\xspot = 0.6$).\\

\begin{figure}
    \centering
    \includegraphics{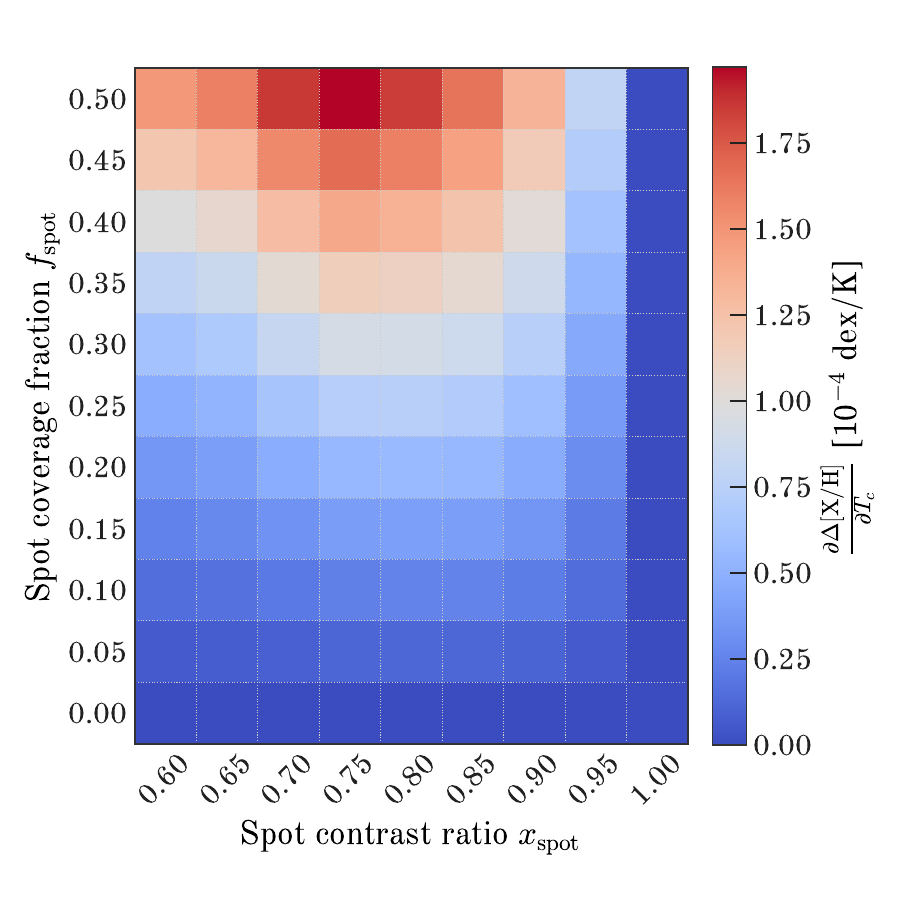}
    \caption{Slope of mean element abundances with respect to condensation temperature ($d\mathrm{[X/H]}/dT_c$) for simulated Solar spectra with varying spot filling fractions \fspot\ and spot contrast ratios \xspot. Observed solar umbrae have contrast ratios $\xspot \approx 0.6$--$0.8$ \citep{Wallace2000} and the Sun has $\fspot \lesssim 5\times10^{-3}$ at solar maximum \citep{Balmaceda2009}; active solar-type stars can reach $\fspot \approx 0.3$ \citep{Cao2022}.\label{fig:solar-tc-slopes}}
\end{figure}

To place these parameters in context: observed solar umbrae have contrast ratios of $\xspot \approx 0.6$--$0.8$ \citep{Wallace2000}, while the Sun's filling fraction rarely exceeds $\fspot \approx 5\times10^{-3}$ even at solar maximum \citep{Balmaceda2009}, making the Solar regime fall in the lower-right corner of Figure~\ref{fig:solar-tc-slopes} where induced \Tc\ slopes are negligibly small. The stars in the \citet{Bedell2018} sample span $\logrhk \approx -5.1$ to $-4.4$, corresponding roughly to $\fspot \lesssim 0.1$ for the majority, with the most active subset approaching the $\fspot \approx 0.3$ plateau \citep{Cao2022}. At these filling fractions, Figure~\ref{fig:solar-tc-slopes} predicts \Tc\ slopes of $\sim 10^{-4}$\,dex\,K$^{-1}$ for contrast ratios typical of observed sunspots, consistent with the amplitudes reported by \citet{Bedell2018}.\\

We first synthesized spectra of the ambient Sun (assuming $T_\mathrm{eff} = 5777\,\mathrm{K}$, $\log{g} = 4.44$, $\mathrm{[M/H] = 0}$, and $v_\mathrm{micro} = 1.0\,\mathrm{km\,s}^{-1}$) using the \texttt{Korg} spectral synthesis code \citep{Wheeler2023,Wheeler2024}. In all syntheses we used the spherical MARCS 1D model atmospheres \citep{Gustafsson2008}, the Solar chemical composition of \citet{Asplund2009}, and the line list from \citet{Melendez2009}. For each $\{\xspot, \fspot\}$ pair we then synthesized a spotted spectrum, holding the surface gravity and metallicity fixed at the ambient values.
We assume the emergent spectrum is a flux-conserving linear combination of the spotted and ambient spectra,
\begin{equation}
    \label{eq:flux-combination}
    F\left(T_\mathrm{eff},\lambda\right) = \fspot{}F\left(\Tspot, \lambda\right) + (1 - \fspot)F\left(\Tamb,\lambda\right)\,,
\end{equation}
where $F$ denotes the emergent flux from the synthesis at a given effective temperature.\\

For each trial of $\{\xspot, \fspot\}$ we treat the rectified (continuum-normalized) emergent spectrum from \texttt{Korg} as if it were a noiseless observation of a Sun-like star, without any instrumental or rotational broadening applied. For every line in our line list we fit the spectrum in a small ($\pm 1$\,\AA) window around the line center using \texttt{Korg.Fit.fit\_spectrum}, with only the transition of interest in the line list, allowing the element abundance to vary until the L-BFGS-B optimization algorithm \citep{lbfgsb} had minimized the $\chi^2$ difference. We kept the stellar parameters fixed at the ambient Solar values throughout, so the inferred line-by-line abundances reflect the combined effect of the spotted spectrum and the (incorrect, unchanged) assumed stellar parameters. We recorded these individual line abundances for all transitions in every $\{\xspot,\fspot\}$ pair.\\

In this and subsequent experiments, we took the mean of line-by-line abundance differences (spot $-$ ambient) for every element, grouping abundances from neutral and ionized transitions together. The uncertainty in each element abundance was taken as the standard error of the mean, i.e., the standard deviation of the line-by-line abundance differences divided by $\sqrt{N}$, where $N$ is the number of transitions. If an element had only one transition then we adopted a minimum uncertainty of 0.01\,dex. When computing the gradient of element abundances with respect to condensation temperature, we used the 50\% condensation temperatures from the Solar system simulation of \citet{Lodders2003}, which we assume to be correct and known perfectly. Following standard practice in the literature, we restrict our fit to only include elements with condensation temperatures exceeding 1000\,K.\\

The \Tc\ slopes we measure from simulated spotted Solar spectra are shown in Figure~\ref{fig:solar-tc-slopes}. All simulations yielded non-negative ($\geq 0$) gradients. As expected, slopes are exactly zero for experiments with no contrast ($\xspot = 1$), regardless of the spot filling factor. At higher contrast ratios (lower $\xspot$) the largest measured slope increases, and experiments with higher spot filling factors show higher slopes with condensation temperature. At face value, this simple spot model suggests that even small spot filling factors and contrast ratios are sufficient to produce an apparent positive trend ($\sim 10^{-4}\,\mathrm{dex\,K}^{-1}$) between mean element abundance and condensation temperature. This is broadly consistent with the magnitude of slopes reported in Figure~\ref{fig:bedell}, with the largest reported value there being $\approx 3.5\times 10^{-4}$\,dex\,K$^{-1}$. However, arguably our \texttt{Korg} simulations could be interpreted as a \emph{lower} limit of \Tc\ given some $\{\xspot,\fspot\}$ pair, as our synthesis did not include molecular transitions, which become increasingly important at cooler temperatures (lower $\xspot$ at any $\fspot$). As a consequence, the differences in inferred line abundances (and \Tc) could be underestimated.\\

\subsection{Analysis of a Solar spot umbral spectrum} \label{sec:observed-solar-spectra}

The simulations described in Section~\ref{sec:simulated-data} are akin to a mixture model, where the data are represented as a linear combination of an ambient Solar spectrum and a spotted one. Here we take a data-driven approach to study the effects that spots might have on inferred abundances. The data consist of two high-quality spectra of the Sun taken with the Kitt Peak Observatory Fourier Transform Spectrograph, a folded Michelson interferometer for obtaining extremely high-resolution observations of the Sun (and other Solar neighborhood bodies). The ambient (quiet) photospheric spectrum we use is taken from the Kitt Peak FTS solar flux atlas of \citet{Wallace1998}, and the sunspot umbral spectrum is from the corresponding FTS umbral atlas of \citet{Wallace2000}. The combined wavelength coverage extends from 3570\,\AA\ to 7405\,\AA, which includes most of the transitions in our adopted line list \citep{Melendez2009}. Both spectra are continuum-rectified with telluric features removed.\\

We take the umbral spot spectrum to represent the extreme case of a highly spotted star. The spotted spectrum is remarkably different from the quiet Solar spectrum \citep[see also][]{Komori2025}, showing substantial molecular absorption features, including H$_2$O \citep{Wallace2000}. In other words, no spectroscopist would claim that these two spectra (the ambient and the umbral) are similar enough for a differential analysis. However, we continued with a differential analysis simply to better understand what the extreme differential effects are when assuming a Solar-like photosphere to analyze a spectrum of an umbral feature. We measured equivalent widths for every transition in our line list in both the ambient and umbral spectra by fitting a Gaussian profile, and converted these to abundances with \texttt{Korg.Fit.ews\_to\_abundances} \citep{Wheeler2023,Wheeler2024}, assuming a Solar photosphere (see adopted parameters in Section~\ref{sec:simulated-data}). We inspected the fits for every transition and found that many lines in the umbral spectrum were unsuitably blended by other features, leading us to discard several transitions, just as we would if we were doing a line-by-line differential analysis.\\

\begin{figure}
    \centering
    \includegraphics[width=1.0\columnwidth]{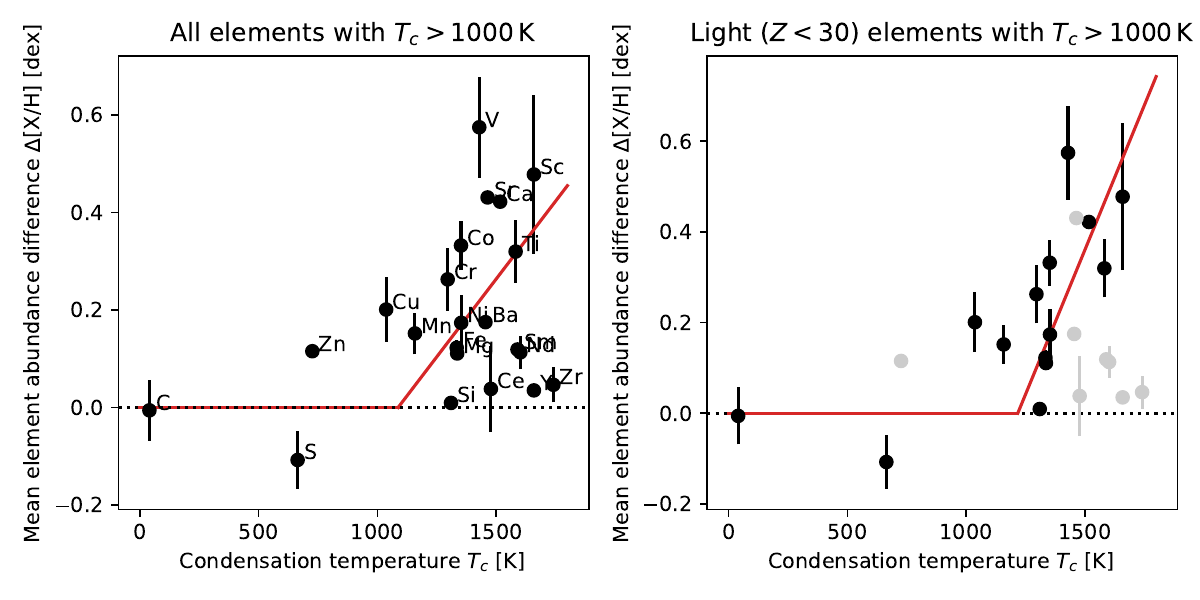}
    \caption{Mean element abundance differences (spot $-$ ambient) inferred from Solar observations as a function of condensation temperature \citep{Lodders2003}. In the right panel, elements with atomic number $Z \geq 30$ are shown as gray points and excluded from the fit.\label{fig:solar-umbral-abundances-wrt-tc}}
\end{figure}

The element abundance differences between the umbral spot spectrum and the ambient spectrum are shown in Figure~\ref{fig:solar-umbral-abundances-wrt-tc}. The mean element abundance differences are substantial, with many elements showing a factor of 3 increase ($>0.5$\,dex) in abundance. Some caveats of this analysis are discussed in Section~\ref{sec:discussion}. One such caveat is the choice of which element abundances should be included in the fit. Some studies chose to vary the minimum condensation temperature necessary for an element to be included (e.g., $T_C > 1000\,\mathrm{K}$ or $T_C > 800\,\mathrm{K}$), and some studies chose to restrict their fit to light elements only (e.g., those with atomic numbers $Z < 30$). For simplicity and consistency with \citet{Bedell2018}, we restrict all our fits to include all elements with $T_C > 1000\,\mathrm{K}$. If we were to exclude heavy element abundances \citep[e.g.,][]{Melendez2009} then this makes the umbral sunspot abundance trend with condensation temperature appear more significant (Figure~\ref{fig:solar-umbral-abundances-wrt-tc} right).\\

\section{Discussion}
\label{sec:discussion}

Stellar activity is already known as a substantial noise source for extreme-precision radial velocities, stellar parameters, and chemical abundances \citep{Crass2021,Wilson2023,Spina2020}. This motivates ongoing work on modeling stellar atmospheres to mitigate the effects of stellar activity. In general, starspots make the photosphere appear cooler, causing lines with low excitation potential (or elements with low FIP) to increase in strength (equivalent width). Transitions that are more sensitive to Zeeman splitting can also appear broader, and therefore have larger equivalent widths. Because magnetic field strength is not normally included as a nuisance parameter when measuring stellar abundances, the larger equivalent width translates directly into a higher inferred abundance. For some elements there is only a limited selection of atomic lines that can be measured in stellar atmospheres, where elements with low FIP have strong resonance and low-excitation lines at optical wavelengths. As a consequence, starspots can bias inferred chemical abundances in subtle ways, with elements that have lower FIP more likely to be affected by stellar activity.\\

More generally, any variation in the conditions of the outer photosphere could translate into a subtle abundance signature because elements are affected to varying degrees. These include unmodeled physical structure such as starspots, as well as inaccuracies in the model itself (e.g., in the electron number densities interpolated from existing grids of model atmospheres). In both cases the problem is the same: the adopted model does not describe the photosphere being observed, and the mismatch propagates into biased stellar parameters and abundances. This is not a new realization. \citet{Ramirez2009} themselves considered the possibility that measured abundance differences might be caused by variations in first ionization potential. While a correlation was found, they discarded this scenario in favor of a causal explanation involving rocky planet formation. \citet{Yu2025} reach a similar conclusion from co-natal stellar pairs, finding that the more refractory-depleted star in each pair is also more magnetically active, and interpreting the correlation as an imprint of planet formation on pre-main-sequence magnetic evolution. We note that the \citet{Yu2025} result is, however, equally consistent with the spot-mimicry mechanism we describe here: if stellar activity causes apparent refractory depletion at any single epoch (as our three experiments demonstrate), then the more active member of a co-natal pair will systematically appear more refractory-depleted, producing precisely the correlation they observe. Distinguishing between these scenarios requires either monitoring the \Tc\ slope of individual active stars across their activity cycles, or identifying elemental outliers --- such as sodium (Na), with a low FIP but only intermediate \Tc\ --- that should respond differently under spot mimicry versus genuine planet formation. The planet-formation hypothesis has open questions related to the short timescale of planet formation relative to the pre-main-sequence phase, and whether any such abundance signature could persist in the photosphere given that young stars are still expected to be (nearly) fully convective \citep[e.g.,][]{Behmard2023a}. Despite these open questions, much of the subsequent literature has continued to interpret abundance patterns with respect to the \citet{Lodders2003} condensation temperature values as evidence of planet formation.\\

In the simulations we performed where spots are added to the Sun, we find that the \Tc\ slope varies smoothly with spot contrast ratio $\xspot$ and spot filling fraction $\fspot$. This is a relatively simple model for spots that does not account for spot size, geometry, latitude distribution, or decay timescales. Nevertheless, the simple parameterisation is sufficient to represent the integrated spectrum from a Sun-like star. With fixed Solar parameters, the computed \Tc\ slopes vary by at least $\sim 10^{-4}\,\mathrm{dex\,K}^{-1}$. In other words, with a simple spot model we can produce \Tc\ slopes that are consistent with what is interpreted in the literature as a diagnostic signature of rocky planet formation or engulfment. \\

Our simulations are intended to test plausibility, not to make a quantitative prediction. They have several shortcomings that make a direct comparison with observed \Tc\ slopes unrealistic: they use 1D LTE model atmospheres, omit molecular opacity and Zeeman splitting, use a two-parameter spot model with no geometry, and hold the stellar parameters fixed at Solar values. Since spots change the emergent spectrum in a way that partly mimics a change in stellar parameters, re-solving for $T_\mathrm{eff}$, $\log{g}$ and $v_\mathrm{micro}$ (e.g., by excitation and ionization balance) would absorb some of the signal. The net effect on the \Tc\ slope is not obvious a priori: a lower $\log{g}$ weakens pressure-sensitive lines of ionized species (predominantly low-FIP refractories), but introduces its own correlated bias between neutral and ionized species. We expect this to reduce but not eliminate the spot-induced trend, and leave a quantitative test to future work. More broadly, nearly every study that has used abundance patterns to infer planet formation or engulfment has relied on 1D LTE photospheres, on the assumption that model deficiencies cancel out in a differential analysis between similar stars. That assumption fails when the two stars differ in their level of activity. For these reasons the empirical evidence (Figure~\ref{fig:bedell}) is more compelling than the simulations, which show only that plausible spot configurations produce \Tc\ slopes of the right sign and amplitude.\\

One might think that the cleanest empirical test of how starspots affect chemical abundances would be to compare disk-integrated Solar spectra taken at peak versus minimum activity. Indeed, time-series data of this kind already exist \citep[e.g., the HARPS-N solar telescope][]{Cegla2019, Dumusque2021}, but the Sun is too inactive for such a test to be informative. While it does vary on an 11-year cycle, the peak activity is very low. Over the Solar cycle, \logrhk\ varies between approximately $-4.85$ at cycle maximum and $-4.97$ at cycle minimum, an amplitude of only $\sim 0.12$\,dex \citep{Mamajek2008}. Even during the strongest solar cycles on record, the monthly-mean sunspot coverage has not exceeded $\fspot \approx 5\times10^{-3}$ \citep{Balmaceda2009}. Placing these levels of activity on Figure~\ref{fig:bedell}, we would not expect to measure any significant difference in \Tc\ slope between Solar observations taken at maximum and minimum activity: Solar neighborhood stars with $\logrhk \approx -5$ have \Tc\ slopes consistent with zero, and a $\Delta\logrhk \approx 0.12$\,dex translates to a change in \Tc\ slope of only $\sim 5 \times 10^{-5}\,\mathrm{dex\,K}^{-1}$, well within the uncertainties \citep{Bedell2018}. For this reason, we do not pursue a Solar-cycle differential analysis here; a meaningful empirical test of the spot--abundance link instead requires more active stars, where the same activity-cycle approach should yield a measurable signal. Promising candidates are moderately active solar-type stars with multi-year archival baselines of high-resolution spectra---instruments such as HARPS and the California Planet Search have operated long enough to capture multiple activity cycles for such stars, and EXPRES is well-suited to this test given its extreme precision and broad spectral coverage.\\

We analyzed an umbral spot spectrum instead, which gives a measure of the maximum abundance difference that might result from the presence of spots. The umbral spectrum is remarkably strange compared to spectra usually encountered by a stellar spectroscopist. Many lines that are weak in the ambient Solar spectrum are completely saturated in the umbral spectrum. Many transitions show evidence of strong magnetic splitting or broadening. There are strong molecular bands, including H$_2$O and C$_2$. Even if this analysis were repeated with non-LTE, it would have to include magnetic splitting of lines and complete molecular line lists.\\

Perhaps the most compelling evidence is the relation between \logrhk\ and the \Tc\ slope of Solar neighborhood stars shown in Figure~\ref{fig:bedell}, using the sample of \citet{Bedell2018}. Stars with higher \logrhk\ systematically show larger \Tc\ slopes. This is consistent with our imperfect simulations of starspots, and with our analysis of the umbral sunspot spectra, which represent our best estimate of the maximum abundance difference due to spots. While the abundance pattern we find between the ambient and umbral Solar spectrum does produce a slope of element abundances with condensation temperature, the relationship between the two quantities only appears linear for the light element abundances ($Z < 30$). If this is true, it is good news: it would suggest that the impact of starspots might be distinguished from any genuine signatures of planet formation and galactic chemical evolution, or that the effect of spots could be simultaneously modeled when inferring chemical abundances. Moreover, some elements could be particularly informative outliers: sodium (Na) is a strong electron donor with a very low first ionization potential and a moderate condensation temperature ($T_c \approx 958$\,K; \citealt{Lodders2003}). Whether the Na abundance proves a useful discriminant between starspots and planet formation will require more realistic modeling of spots that includes non-LTE effects and magnetic fields.\\

Until such models become available, a na\"ive approach to mitigate the dependence on starspots might simply be to restrict the sample to stars with $\logrhk \approx -5$, or to restrict the analysis to transitions of comparably high excitation potential. For example, in our Solar simulations the transitions with excitation potentials of about 5\,eV had the smallest abundance differences. An alternative approach might be to perform a first-order correction to the \Tc\ slope based on the measured level of stellar activity (through \logrhk). However, we stress that controlling for activity statistically is difficult. \logrhk\ is a chromospheric proxy, not a direct measure of spot coverage or contrast. Its relationship to spot properties changes between the plage-dominated regime of Sun-like activity and the spot-dominated regime of more active stars, and it depends on inclination and on the photospheric correction applied for $T_\mathrm{eff}$. Activity is also correlated with age, which is itself correlated with \Tc\ slope through Galactic chemical evolution, and activity varies in time, so a single (or co-added) \logrhk\ measurement need not represent the spot configuration at the epochs used for abundance analysis. A regression on \logrhk\ is therefore unlikely to remove the activity signal completely, and residual correlations should be expected. For stars where a single-epoch correction is not enough -- for example, more active stars, or stars where activity is the dominant systematic -- it may be possible to recover reliable abundances by measuring them at several phases of the activity cycle and either marginalizing over, or explicitly modeling, the resulting variation. Although we do not demonstrate this directly, our results imply that the spread in \Tc\ slope across an activity cycle is itself a measurable quantity that could be used to bound the activity-induced contribution to a star's mean abundance pattern, allowing the underlying photospheric composition to be disentangled from the time-varying spot signal. Such an approach would convert the problem from one that is invisible at single-epoch precision into one that can be calibrated out, potentially extending precision-abundance work to the active stars where it is currently most unreliable. It is clear that detailed abundance analyses that do not explicitly account for stellar activity can be misled by active stars, and that stellar activity may be imposing a lower limit on the intrinsic abundance dispersion of open clusters, thereby setting a noise floor for chemical tagging. Specific targets well-suited to such a test include moderately active solar-type stars with $\logrhk \gtrsim -4.7$ that have been observed across multiple activity cycles at high spectral resolution and signal-to-noise --- for instance, stars in the \citet{Bedell2018} sample with the highest measured $\logrhk$ values, or active planet-hosting stars with long HARPS or California Planet Search baselines.

\section{Conclusions}\label{sec:conclusions}

We have shown that stellar activity can produce abundance patterns that mimic the amplitude of reported \Tc\ trends in solar-type stars. In the HARPS solar-twin sample of \citet{Bedell2018}, the \Tc\ slope correlates with the chromospheric activity indicator \logrhk, with the least active stars showing \Tc\ slopes consistent with zero. An independent line of evidence consistent with this picture comes from \citet{Yu2025}, who find that within co-natal co-moving pairs the more refractory-depleted star is also more magnetically active, with the differential trend strengthening with \Tc. While they interpret this as an imprint of planet formation, this correlation is also a natural and quantitatively expected consequence of spot-induced spectroscopic biases. In simulations of the spotted Sun, plausible combinations of spot filling fraction and contrast can reproduce \Tc\ slopes that are commonly interpreted in the literature as evidence of planet formation or engulfment. And in a line-by-line differential analysis of a disk-integrated Solar spectrum against a sunspot umbral spectrum, we find that refractory elements show differences of up to $\sim 0.5$\,dex. These three experiments highlight the often under-appreciated importance of stellar activity in precision spectroscopy.\\

The causal argument that a star's \Tc\ slope is a diagnostic of rocky planet formation principally depends on the fact that we know the Sun has rocky planets, which might explain why the Sun's refractory element abundances differ from other Solar-like stars.
However, while we don't know which stars do or do not host rocky exoplanets, we do know that stars have spots. Activity -- at some level -- is a universal feature of the stars in question, but rocky planets are not. And even though the Sun is chromospherically quiet, the maximum $\Delta\logrhk$ across the Solar cycle could translate into a change in \Tc\ slope of $\approx 5 \times 10^{-5}\,$dex\,K$^{-1}$, a signal that is often interpreted as evidence of rocky planet formation. Any inferences of a planet-formation or engulfment signature from a \Tc\ slope should at least demonstrate that the target is chromospherically quiet, recognizing that this is a necessary but not sufficient condition.

\begin{acknowledgements}
The Flatiron Institute is a division of the Simons Foundation. 
This work made use of HARPS spectra obtained at the ESO 3.6\,m telescope at La Silla Observatory under ESO programme IDs 072.C-0488, 074.C-0364, 075.C-0202, 075.C-0332, 076.C-0155, 077.C-0364, 088.C-0323, 089.C-0415, 089.C-0732, 090.C-0421, 091.C-0034, 091.C-0936, 092.C-0721, 093.C-0409, 095.C-0551, 096.C-0499, 097.C-0090, 097.C-0571, 097.C-0948, 183.C-0972, 183.D-0729, 185.D-0056, 188.C-0265, 192.C-0224, 192.C-0852, 289.C-5053, 292.C-5004, and 60.A-9036, as compiled by \citet{Bedell2018}.
\end{acknowledgements}

\bibliography{main}{}
\bibliographystyle{aasjournal}

\end{document}